\documentstyle[11pt,epsfig]{article}
\def\be{\begin{eqnarray}}\def\ba{\begin{eqnarray}}
\def\ee{\end{eqnarray}}\def\ea{\end{eqnarray}}
\def\ben{\begin{enumerate}}\def\bitem{\begin{itemize}}
\def\een{\end{enumerate}}\def\eitem{\end{itemize}}

\def\la{\langle}\def\ra{\rangle}

\def\no{\nonumber\\}

\def\roughly#1{\mathrel{\raise.3ex\hbox{$#1$\kern-.75em%
\lower1ex\hbox{$\sim$}}}}

\def\A0{A_0}
\def\bq{\begin{equation}}
\def\eq{\end{equation}}

\begin{document}
\begin{titlepage}

\begin{flushright}
SNU-TP 01-005
\end{flushright}

\begin{center}

\vskip 3.5cm
{\Large Meson Exchange Effect on  Color Superconductivity  }

\vskip 2.3cm
    {\large Youngman Kim, Young-Ho Song and Dong-Pil Min }
\vskip 0.5cm

 {\it  Department of Phyiscs, Seoul National University, Seoul 
151-742, Korea}

\end{center}

\vskip 1cm

\centerline{(\today)}
 \vskip 2cm

\centerline{\bf Abstract}
 \vskip 0.5cm

We investigate the effects of pion and gluon exchanges on the formation of two-flavor color
superconductivity  at moderate density, $\mu <1 GeV$. 
The chiral quark model proposed by Manohar and Georgi containing pions
as well as gluons is employed to show that
 the pion exchange reduces substantially the value of the superconducting gap gotten with 
the gluon exchange only. 
It turns out that the pion exchanges produce a repulsion between
quark-quark pair in a spin and isospin singlet state.
 We suggest that the phase consisiting of pions, gluons and quarks is one of the candidates
of in-medium QCD phase at moderate density.
\end{titlepage}
\newpage

\section{Introduction}

Due to the attractive interactions between two quarks from
one-gluon exchange or 't Hooft interaction induced by instanton, 
there is a tendency toward spontaneous breaking of (color) gauge symmetry forming
color supercondutivity and especially  color-flavor-locking(CFL)
 \cite{csc}\cite{arw}. The magnitude of the superconducting gap $\Delta$
is estimated by using models whose parameters are chosen to reproduce reasonable 
zero density physics \cite{csc}\cite{arw}
 or from the perturbative one-gluon exchange calculations which
are valid at asymptotically high density\cite{tf}\cite{son}.
Recently, it is shown that at intermediate densities $\mu\sim 500 MeV$
, large enough for the system to be in the
quark phase but small enough to support nonperturbative interactions,
 Cooper pairs
with nonzero total momentum are favored leading to gaps which vary in space\cite{sbmit}.
For recent reviews on the color superconductivity in quark matter, see Ref. \cite{rwh}.

On the other hand, the dense nuclear matter has been
successfully described in terms of the hadrons, which will be called {\it hadronic} phase. 
It was already conceived that the
boson condensation, particularly kaon condensation, can be
produced at the relatively low density, about 3-4 times of the
normal nuclear matter density. Indeed, the hadrons as
the quasi-particles are subdued to the pattern of BR (Brown-Rho)
scaling\cite{br}, whose relation to the results based on the conventional
hadronic many-body calculation\cite{rapp} is investigated in Ref.\cite{krbr}.

Between the hadronic and quark phases, we could expect a phase
whose relevant degrees of freedom are mesons as Goldstone bosons of spontaneous
breaking of chiral symmetry by non-zero 
value of chiral order parameter $\la\bar\psi\psi\ra$ and gluons due to
lack of confinement as well as  constituent (quasi) quarks\cite{mg}, 
and we will call this phase as the
CQ(constituent or chiral quark) phase. Phase diagram for QCD involving hadronic(nuclear) matter
and quark matter(2SC, CFL) is conjectured and depicted in \cite{unlock}.
We propose that the CQ phase is one of the candidates
of in-medium QCD phase at moderate density.
A recent study on
baryons in matter-free space reveals that the ground states and excitation spectra of
light and strange baryons 
 is governed by the Goldstone boson exchange(GBE) as well as 
the harmonic confinement potential 
between constituent quarks (quasi-quarks) \cite{gloz}.
Based on the decoupling of vector interactions at high density,
it is argued in Ref. \cite{gmc} that quasi-particle picture involving
(quasi) quarks becomes more appropriate at higher densities as chiral restoration
is approached. 

The goal of this work is to study two-flavor color superconductivity (2SC) in the CQ phase
and therefore to investigate a phase boundary of hadronic matter and quark matter 
(especially 2SC).
We calculate the gap of two-flavor color superconductivity
in the CQ phase  using the chiral quark model proposed by
Manohar and Georgi\cite{mg}. Note that the model \cite{mg} is valid around the energy scale
 $E_{CQ}$
given in matter-free space $\Lambda_{QCD}<E_{CQ}<\Lambda_{\chi SB}\sim 1 GeV $. We expect, however,
that the model is relevent to describe QCD at moderate density since we can argue that $\mu \sim
E_{CQ}$ based on the renormalization group analysis at finite density\cite{man}.

 In section 2, we introduce the chiral
quark model and derive a gap equation for the Cooper pair
from one gluon and one pion exchange. To see the physics concealed
in the gap equation economically, we adopt contact
four-Fermi interactions, which are assumed to reproduce the physics of one
gluon and one pion exchange, and solve the gap equation in section
3. Discussions are presented in section 4.

\section{Schwinger-Dyson equation in the Chiral Quark Model}

The chiral quark  model is defined between $\Lambda_{QCD}\sim 200
MeV$ and $\Lambda_{\chi SB}\sim 1 GeV$ and therefore contains
mesons as Goldstone bosons of $SU(3)_L\times SU(3)_R$ spontaneous
chiral symmetry breaking. And gluons as gauge bosons of
$SU(3)_{color}$ should be also present in this phase of
deconfinement. 
The chiral quark lagrangian is given by 
\ba {\mathcal
L}&=&{\bar\psi}(i {\not\! D}+{\not\! V})\psi + g_{A}{\bar
\psi}{\not\! A}\gamma_5\psi
               - m {\bar \psi}\psi + \mu{\bar\psi}{\gamma}^0\psi\no
               &&+ \frac{1}{4}f_\pi^2 
tr(\partial^{\mu}\Sigma^{\dag}\partial_{\mu}\Sigma)
               - \frac{1}{2}tr(F^{\mu\nu}F_{\mu\nu}) + \ldots \no
\no
\label{lag}
\ea
where
\ba
D_\mu&=&\partial_\mu+i g G_\mu,~~~ G_\mu=G_\mu^a T^a, \no
V_\mu&=&{i \over 
2}(\xi^{\dagger}\partial_\mu\xi+\xi\partial_\mu\xi^{\dagger}),\no
A_\mu&=&{i \over 
2}(\xi^{\dagger}\partial_\mu\xi-\xi\partial_\mu\xi^{\dagger}),\no
\xi&=&e^{(i \Pi/f_\pi)},~~~ f_\pi\simeq 93~ MeV,~
\Sigma=\xi\xi\nonumber
\ea
and
\ba
\Pi={1 \over\sqrt{2}}\left( \begin{array}{ccc}
                            \sqrt{1 \over 2}\pi^0+\sqrt{1 \over 6}\eta& 
\pi^+& K^+ \\
                            \pi^-&-\sqrt{1\over 2}\pi^0+\sqrt{1\over 
6}\eta& K^0 \\
                            K^- & \bar{K}^0& -{2 \over \sqrt{6}}\eta
                            \end{array} \right).
\ea
Note that in the chiral quark model, {\it we can establish
a power counting argument to deal with non-renormalizable terms and to 
neglect
internal gluon lines} \cite{mg}.

Since the chiral quark model is valid below the chiral symmetry
breaking scale, the value of the chiral order parameter $\la\bar\psi\psi\ra$ is not
zero. We expect, however, the value of $\la\bar\psi\psi\ra$ in CQ
phase  to be quite small compared to that in the matter-free
space.
To see the effects of GBE on color superconductivity, we consider
one pion exchange as well as one gluon exchange with $N_f=2$. As
it is described in the conjectured phase diagram for QCD at zero
temperature ~\cite{abr}, when chemical potential exceeds the
constituent strange quark mass the more relevant phase will be
$2SC+s$ in which we have $ss$ condensates as well as Cooper pairs
composed of $u$ and $d$ quarks. We can, however, neglect the
difference between the $2SC$ and $2SC+s$ since the $ss$ condensate
is expected to be small \cite{abr}. In this work, we ignore $ss$
condensate and concentrate on $2SC$. Since the value of chiral
order parameter is assumed to be small, we may take the light
chiral quark masses to be negligible compared to chemical
potential. There are several works on the possibility of mixed
phase of both chiral condensate and color
superconductivity\cite{ms}.

In this work, we use the Nambu-Gorkov formalism with
\begin{displaymath}
\Psi \equiv \left( \begin{array}{c} \psi \\ {\bar\psi}^T \end{array} 
\right)
             \equiv\left( \begin{array}{c} \psi \\ \psi_c \end{array} 
\right).
\end{displaymath}

The inverse quark propagator
\footnote{Note that in the 2SC phase Cooper pairs of quarks cannot be color singlets and
pick a color direction, the 3 direction for example. Therefore the color 3 quarks remain ungapped.
 } is
\be
S^{-1}(q)=\left(\begin{array}{cc}{\not\! q}+{\mu}\gamma_0-m&\bar \Delta 
\\
                                 \Delta      & ({\not\! 
q}-{\mu}\gamma_0+m)^T \end{array} \right)
\ee
where $\bar\Delta =\gamma_0\Delta^{\dag}\gamma_0$.
To leading order in the perturbative expansion, the quark-gluon vertex 
matrix is
$-i g \Gamma^a_{\mu}$ and the quark-pion vertex matrix is
given by $- (g_A / f_\pi)\Gamma^i(q)$.
\ba
\Gamma^a_{\mu}&=& \left( \begin{array}{cc}T^a{\gamma}^{\mu} &0 \\
                   0  &-(T^a{\gamma}^{\mu})^T \end{array} \right) \no
\Gamma^i(q)&=&  \left( \begin{array}{cc}\frac{\tau^i}{2}{\not\! 
q}\gamma_5 &0 \\
                 0  &-(\frac{\tau^i}{2}{\not\! q}\gamma_5)^T 
\end{array} \right)
\ea
The
gluon propagator in the hard dense loops (HDL) approximation is given 
by
\be
D_{\mu\nu}(q)=\frac{P_{\mu\nu}^T}{q^2-G}+\frac{P_{\mu\nu}^L}{q^2-F}-\xi\frac{q_\mu 
q_\nu}{q^4}
\ee
where
\ba
P_{ij}^T&=&\delta_{ij}-{\hat q}_i{\hat q}_j ,~ P_{00}^T=P_{0i}^T=0 \no
P_{\mu\nu}^L&=&-g_{\mu\nu}+\frac{q_\mu q_\nu}{q^2}-P_{\mu\nu}^T  
\nonumber
\ea
For $q_0\ll{\vec q}\rightarrow 0 $ and to leading order in
perturbation theory we have,
\ba
F=2 m^2,~ G={\pi \over 2}m^2{q_0 \over |{\vec q}|},\nonumber
\ea
with $m^2=N_f g^2\mu^2/(4\pi^2) $.
The pion propagator is $D(k)=1/k^2$.

The Schwinger-Dyson equation for gap matrix $\Delta$ becomes,
\ba
\Delta(k)&=&i g^2 \int\frac{d^4 q}{(2\pi)^4}(-T^a\gamma^{\mu})^T 
S_{21}(q)
              (T^b\gamma^{\nu}) D_{\mu\nu}(q-k) \no
          &&+ i(\frac{g_A}{f_{\pi}})^2 \int\frac{d^4 q}{(2\pi)^4}
             (-\frac{\tau^i}{2}({\not\! q}-{\not\! k})\gamma_5)^T 
S_{21}(q)
             (\frac{\tau^i}{2}({\not\! k}-{\not\! 
q})\gamma_5)\delta^{ij} D(q-k) \nonumber
\ea
Let us take the form of the gap matrix as \cite{tf}
\be
\Delta^{ab}_{ij}(q)=(\lambda_2)^{ab}(\tau_2)_{ij}C\gamma_5
                    [\Delta_1(q_0)\frac{1+{\vec \alpha}\cdot{\hat 
q}}{2}
                     +\Delta_2(q_0)\frac{1-{\vec \alpha}\cdot{\hat 
q}}{2}]
\ee
where ${\vec \alpha}=\gamma_0 {\vec \gamma} $.
By inverting the inverse quark propagator matrix  $S^{-1}(q) $, we 
obtain
the 21-component of $S(q)$
\ba
S_{21}(q)=-(\lambda_2\tau_2C\gamma_5)
          (\frac{\Delta_1}{q_0^2-(|{\vec q}|-\mu)^2-\Delta_1^2} 
\Lambda_{-}
           +\frac{\Delta_2}{q_0^2-(|{\vec 
q}|+\mu)^2-\Delta_2^2}\Lambda_{+})\nonumber
\ea
where $\Lambda_{+}=(1+{\vec \alpha}\cdot{\hat q})/2 $ and
$\Lambda_{-}=(1-{\vec \alpha}\cdot{\hat q})/2 $.
For color and isospin, we can use the following relation,
\be
\frac{1}{4}(\lambda_a)^T\lambda_2\lambda_a=-\frac{N+1}{2N}\lambda_2
                                  =-\frac{2}{3}\lambda_2 ,~(N_c =3)
\ee
where we have used
$(\lambda^a)_{ij}(\lambda^a)_{kl}=-2/N\delta_{ij}\delta_{kl}
                    +2\delta_{il}\delta_{jk}$.
For isospin $N_f=2$, we use
$\frac{1}{4}(\tau_i)^T\tau_2\tau_i=-\frac{3}{4}\tau_2$.
Then the gap equation becomes
\ba
\Delta(k)&=&
           \lambda_2\tau_2 C\gamma_5 (\Delta_1(k_0) 
\Lambda_{+}+\Delta_2(k_0)\Lambda_{-})
           \no
 &=& i g^2(+\frac{2}{3}\lambda_2\tau_2 C\gamma_5)
           \int\frac{d^4 q}{(2\pi)^4}
                \gamma_{\mu}(\frac{\Delta_1}
                  {q_0^2-(|{\vec q}|-\mu)^2-\Delta_1^2}\Lambda_{-}\no
            && +\frac{\Delta_2}
                  {q_0^2-(|{\vec q}|+\mu)^2-\Delta_2^2}\Lambda_{+} )
                \gamma_{\nu}D^{\mu\nu}(q-k) \no
          &&+i (\frac{g_A}{f_{\pi}})^2
             (+\frac{3}{4}\lambda_2\tau_2C\gamma_5)
            \int\frac{d^4 q}{(2\pi)^4}({\not\! q}-{\not\! k})\gamma_5
            (\frac{\Delta_1}
                  {q_0^2-(|{\vec q}|-\mu)^2-\Delta_1^2}\Lambda_{-}\no
            && +\frac{\Delta_2}
                  {q_0^2-(|{\vec q}|+\mu)^2-\Delta_2^2}\Lambda_{+} 
)(\not\! q-\not\! k)
\gamma_5 D(q-k)
\ea
Following the same procedure shown in Ref. \cite{tf}, we get the gap 
equation,
\ba
\Delta(k_0) &=& {g^2 \over 18\pi^2}\int^{\infty}_0 dq_0
           \frac{\Delta(q_0)}{\sqrt{q_0^2+\Delta^2(q_0)}}ln({b\mu \over 
|k_0-q_0|}) \no
         &&-{3 \mu^2 \over 16 \pi^2}{g_A^2 \over 
f_\pi^2}\int^{\infty}_0 dq_0
           \frac{\Delta(q_0)}{\sqrt{q_0^2+\Delta^2(q_0)}}
           \frac{(k_0-q_0)^2}{2\mu^2} ln (\frac{4\mu^2}{|k_0-q_0|^2}).
\label{gapeq} \ea 
From (\ref{gapeq}) we can see that one pion
exchange gives repulsive interaction in the quark-quark potential
and therefore the magnitude of the gap will be reduced in the
presence of the GBE. Note that the quark pairs are in isosinglets
and spin singlet state. The reason that one pion exchange(OPE)
gives repulsive potential in the quark-quark potential can be
understood from the similarity with nucleon-nucleon
potential from OPE in the same isosinglet and spin singlet
channel, which is repulsive\cite{ew} 
\ba V_\pi
(S=I=0)=\frac{3f^2}{4\pi}\frac{e^{-m_\pi r}}{r}.
 \ea
We note here that it is not simple to solve the gap equation (\ref{gapeq}) analytically
 even with the approximation given in \cite{son}. 

\section{A Toy Solution}

The property of the gap equation (\ref{gapeq}) can be understood
by a toy solution in which we set forth the following four Fermi
interactions mimicking  the quark-quark interaction in the chiral
quark model. Note that with a NJL type interaction our gap will
have different dependence on the coupling constant
$\Delta\sim\exp(-1/g^2)$ from the one with one-gluon exchange
$\Delta\sim\exp(-1/g)$\cite{son}. We can see, however, the effects
of the GBE on the gap explicitly in a simple manner with the NJL
type interaction.
\be
{\mathcal L}_{int} = -G \int d^4 x ({\bar \psi}\lambda^a \gamma^{\mu} 
\psi)
                                  ({\bar \psi}\lambda^a \gamma_{\mu} 
\psi)
                  +G_{\pi} \int d^4 x({\bar \psi}\vec\tau\gamma_5 
\psi)\cdot
({\bar \psi}\vec\tau\gamma_5 \psi)\label{ml}
\ee
where the first term represents gluon induced interaction,
and the second one is pion induced interaction. Here $\lambda^a$ is a 
generator of $SU(3)$
color group and $\vec\tau$ is for $SU(2)$ isospin group.
Note that we choose positive sign in front of pion induced term 
considering repulsive OPE potential (\ref{gapeq})
and take $G$ and $G_\pi$ to be positive. Since pion is almost massless,
it is not simple to put the one-pion exchange into the four-Fermi form,
but here we expect that the four-Fermi interaction with $G_\pi$ in (\ref{ml})
could at least mimic the {\it repulsive} nature of the one-pion exchange calculated in
section 2.

Now it is straightforward to obtain a gap equation with the interacting Lagranigan (\ref{ml})
and we get
\ba
\Delta
&=&(\frac{2 G \mu^2}{3 \pi^2}-\frac{3 G_\pi \mu^2}{16 \pi^2})
                     \int d q_0 \frac{\Delta}{\sqrt{q_0^2+\Delta^2}} 
\label{sgap}
\ea
Note that this gap equation can be traced from the 
equation (\ref{gapeq}) by
removing propagator dependence, {\it i.e.}
by neglecting logarithmic term in (\ref{gapeq}).
For the sake of simplicity, we introduce cut-off $\Lambda$ and take 
$\Lambda\sim\mu$.
Since the gap $\Delta$ is constant with the interaction (\ref{ml}), we 
get a solution easily
\ba
\Delta &=& 2\mu \exp {(-{1/ N})}
\ea
with $N \equiv \frac{2 G \mu^2}{3 \pi^2}-\frac{3 G_\pi \mu^2}{16 
\pi^2}$.
Now we can see that due to the repulsive nature of pion exchange 
potential,
the value of the gap will be reduced by a factor $\sim\exp 
[-1/(G_\pi\mu^2) ]$.

To estimate the value of the gap,
we take $G\approx 5.8GeV^{-2}$ and
\ba
G_\pi=c\frac{g_A^2}{f_\pi^2}
\ea
where $c$ is a constant. The value of $c$ could be determined by fully 
integrating out
pions to obtain effective action containing four-Fermi interactions, 
but we take it as
a free parameter  and we shall use $g_A= 0.75 $ and $f_{\pi}= 93 MeV$
given in free space for numerical estimates.

\begin{figure}
\centerline{\epsfig{file=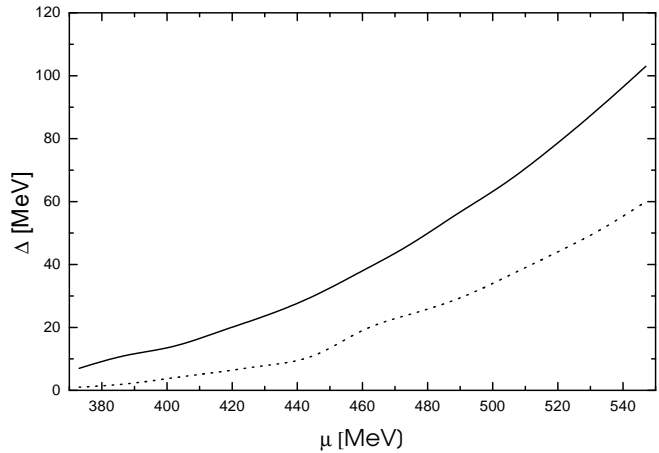,width=10.2cm}}
\caption{\small $\Delta$ as a function of chemical potential. Dotted 
line represent
$\Delta$ with both one gluon and one pion exchange and
solid line is the resulting gap from only one-gluon exchange 
contribution. }\label{cscp}
\end{figure}

Taking  $G\approx 5.8GeV^{-2}$ to reproduce the results 
from NJL model
calculations with only single gloun exchange effects
and choosing $c=0.05$($G_\pi\approx 3.3 GeV^{-2}$), 
we get the results depicted in Fig. \ref{cscp}. 

We note here that when the value of $G_\pi$ is big enough to make $N$ 
to be negative,
we see from (\ref{sgap}) that there is a possibility to have
no color superconducting gap at all.

\section{Discussion}

In this work, we employ
 the chiral quark model to study the effect of one pion exchange
on the two flavor color superconductivity. We found that OPE reduces the
value of the gap $\Delta$, though it is difficult to be quantitative as to how 
much the gap will be reduced
since we have no definte information on the value of $G_\pi$.
Our work will have several important consequences :

$\bullet$
At moderate density where CQ phase is plausible,
  pion as a Goldstone boson
of spontaneous breaking of chiral symmetry by quark condensate $\la 
\bar \psi \psi \ra$ could do important role in the
physics of color superconductivity, implying that at moderate density
there could be some other interactions as well as one gluon exchange
which have to be taken into account. Besides the GBE mode considered in this work,
light vector mesons could come in as indicated by BR scaling \cite{br} and 
the ``vector manifestation''\cite{hy} in which the chiral symmetry is restored
at the critical point by the massless degenerate pion and (longitudinal) $\rho$-meson as the
chiral partner.
In the case of two-nucleon states of total isospin $0$ and spin $0$,
 the $\rho$ exchange potential is given by \cite{ew}
\ba
V_\rho (S=I=0)=3[-\frac{g_\rho^2}{4\pi} +2\frac{f_\rho^2}{4\pi} ]\frac{e^{-m_\rho r}}{r}
\ea
where $f_\rho=(g_\rho +g_t)(m_\rho/2m_N)$. When we take the value of the couplings determined in
free space, we can easily see that the $\rho$ exchange potential is repulsive.
Therefore we expect {\it naively} that light vector mesons also produce
 repulsion between quark-quark pair we are considering. 
We note, however, that the potential becomes attractive when neglecting 
tensor coupling ($g_t=0$) and that density dependence of $g_\rho$ and $g_t$
should be properly taken into account. 
In the vector manifestation, we should also keep it in mind that the transverse
 $\rho$ is decoupled from the vector current.


$\bullet$
The effects of diquark condensates on the cooling of compact stars
is investigated by several authors, see for example ~\cite{bt}.
However it is expected that the quark matter core in neutron star
is {\it inert} as far as cooling is concerned since the quasi-quarks 
contribution to
the specific heat and neutrino emissivity is suppressed by a factor
$\sim \exp (-\Delta /T )$~\cite{alford}.
However, when GBE reduces the gap substantially, we expect
that quark matter core could play important role in cooling of neutron star
and the formation of neutron star~\cite{cr}.

$\bullet$
Extending our work to three-flavor QCD will be quite interesting.
For example, flavor non-singlet nature of the GBE may induce mixing between
$LL$ and $RR$ condensates which is not possible with only gluon exchange interactions 
and  
reduction in the value of the gap will affect meson masses in CFL phase \cite{genm}
and the physics 
of kaon condensation in quark matter ~\cite{sch}, which will be reported elsewhere.

\vskip 1cm
{\bf Acknowledgements}

We would like to thank Mannque Rho and Deog Ki Hong for discussion.
This work is supported partly by the BK21 project of the Ministry of Education,
 by KOSEF Grant 1999-2-111-005-5 and KRF Grant 1999-015-DI0023.


\end{document}